# Second Law based definition of passivity/activity of devices

Kyle Sundqvist [1], David K. Ferry [2] and Laszlo B. Kish [3†]

[1] *Department of Physics, San Diego State University, 5500 Campanile Drive, San Diego, CA 92182-1233 USA*

[2] *School of Electrical, Computer, and Energy Engineering and Center for Solid State Electronic Research, Arizona State University, Tempe, AZ 85287-5706, USA*

[3] *Department of Electrical and Computer Engineering, Texas A&M University, TAMUS 3128, College Station, TX 77843-3128*

**Abstract.** Recently, our efforts to clarify the old question, if a memristor is a passive or active device [1], triggered debates between engineers, who have had advanced definitions of passivity/activity of devices, and physicists with significantly different views about this seemingly simple question. This debate triggered our efforts to test the well-known engineering concepts about passivity/activity in a deeper way, challenging them by statistical physics. It is shown that the advanced engineering definition of passivity/activity of devices is self-contradictory when a thermodynamical system executing Johnson-Nyquist noise is present. A new, statistical physical, self-consistent definition based on the Second Law of Thermodynamics is introduced. It is also shown that, in a system with uniform temperature distribution, any rectifier circuitry that can rectify thermal noise must contain an active circuit element, according to both the engineering and statistical physical definitions.

## 1. Introduction

A simplified engineering mathematical model about a complex physical system can provide mathematically exact statements. However there is a danger that the model is unphysical when it misses capturing an essential part of the physics of the problem. In such cases, the model may yield incorrect conclusions.

Recently, our efforts to clarify the old question: "Is a memristor a passive or active device [1]?" triggered debates between engineers, who have had advanced definitions of passivity/activity of devices, and physicists with significantly different views about this seemingly simple question. These debates motivated our efforts to test the well-known engineering concepts about passivity/activity in a deeper way, by challenging them with statistical physics.

In [2], the Authors present a survey of commonly used engineering classifications of passivity/activity of electronic devices. Then they introduce the most advanced, *so-called-thermodynamic* definition that does not have the weakness of the former (commonly used) ones. However, we have found that even this most advanced engineering definition fails in certain physical/practical cases. Under certain conditions, it miss-classifies well-accepted passive devices as active, especially when temperature differences or specific nonlinearities are present.

Because of the failures of the established engineering concepts, in this paper we introduce a deeper definition of passivity/activity that is based on statistical physics. The Second Law of thermodynamics makes these definitions possible without having particular knowledge about the structure, circuitry, or other fine details of the device. The new, statistical physical definitions are robust against those challenges where the commonly used engineering definitions fail according to the analysis in [2]. Moreover our statistical physical definitions always classify a resistor as a passive device even when the most advanced engineering definitions fail to do so.

---

† Corresponding Author

Below, we first outline the most advanced engineering, *so-called-thermodynamic*, definition [2] of passivity/activity and, in Section 2, we show two thermodynamic situations where it fails. In Section 3, we introduce our new, *statistical-physical* definitions. Finally in Section 4, we show that any rectifier circuitry that can rectify thermal noise must contain at least one active circuit element.

For the sake of simplicity, we restrict the analysis to time-invariant systems and 1-*port*, that is, to 2-*contact* devices. The engineering approach [2] uses the "available energy" $E_A(x)$ served by the device which is

$$E_A(x) = \sup_{t \geq 0}^{(x)} \left[ \int_0^t -u(\tau) i(\tau) \mathrm{d}\tau \right] , \tag{1}$$

where $u$ and $i$ are voltage and current, respectively; the given $x$ represents the initial conditions at time $t = 0$ (that is, all the initial voltages and currents in the system/circuitry between the two contacts); $\sup_{t \geq 0}^{(x)}$ indicates that the upper limit (supremum) is taken over all $t \geq 0$ long time intervals and all physically possible $\{u(t), i(t)\}$ trajectories. The earlier definition would say the device is claimed to be *strongly passive* from a "thermodynamic" point of view [2] if:

i)  $E_A(x)$ is finite for all physically allowed initial states $x$.

ii) There exists a specific initial state $x_0$ (so-called relaxed state), where $E_A(x) = 0$ .

We acknowledge the intent of the Authors of [2] to include thermodynamics in the definition of passivity/activity and the efforts to choose their wording very carefully. They talk about "physically possible" processes, and describe a quiescent "relaxed state" (probably an attempt to account for thermal equilibrium). However, from a fundamental physics point of view, the name, "*thermodynamic*", for the above engineering definition is obviously incorrect because *no thermodynamical measure*, such as *temperature*, *heat* or *entropy* appears in the above formula. In the next section, we point out two practical situations where even the above advanced engineering definition gives the wrong conclusion.

## 2. Examples where the engineering definition fails

The method we use to crack the engineering definition of activity and passivity is based on the Fluctuation-Dissipation Theory (FDT) of statistical physics. In the classical physical limit, the FDT for a passive electrical impedance $Z(f)$ interrelates the power density spectrum $S_u(f)$ of the thermal noise voltage generated by the impedance with the real part $\mathrm{Re}[Z(f)]$ of that impedance [3]:

$$S_u(f) = 4kT \, \mathrm{Re}[Z(f)] , \tag{2}$$

where Equation 2 is the Johnson-Nyquist formula; $k$ is the Boltzmann constant; and $T$ is the absolute temperature of the free-standing impedance in thermal equilibrium. Even though there are current discussions [3,4] about the FDT's prediction in the quantum region (at very low temperatures and/or very high frequencies), in the classical physical limit Equation 2 is verified by both careful experiments and everyday electrical engineering practice.

## 2.1 Resistors at different temperatures

A direct, well-known consequence [5,6] of Equation 1 is utilized in practice even in secure communications, that a circuit of two parallel-connected resistors with different temperatures will execute a non-zero mean power flow from the warmer resistor to the colder one. Let us suppose that, at our test of passivity/activity by means of the engineering definition above, the *device-in-question* is the $R_1$ resistor at $T_1$ temperature in Figure 1 and the $T_1$ and $T_2$ temperatures are different and stabilized by an active temperature regulation. In the frequency band $\Delta f$, the mean power flow from resistor $R_1$ to resistor $R_2$ is given as [5,6]:

$$\langle P_{12} \rangle = 4k(T_1 - T_2)\Delta f \frac{R_1 R_2}{(R_1 + R_2)^2} \ . \tag{3}$$

If the load resistor $R_2$ is colder than $R_1$, that is $T_2 < T_1$, then according to Equation (3):

$$0 < \langle P_{12} \rangle = \lim_{\substack{t \to \infty \\ t > 0}} \frac{E_{12}}{t} \ , \tag{4}$$

thus

$$\lim_{\substack{t \to \infty \\ t > 0}} E_{12} = \infty \ , \tag{5}$$

which, according Equation 1 and the related engineering definitions mean that $R_1$ is an active device contrary of the common understanding that linear resistors are passive devices. On the other hand, when the temperatures are interrelated in the opposite way, $T_1 < T_2$, then the $R_1$ is classified as a passive device. It is clear that the engineering definition based on Equation 1 is self-contradictory for classifying the activity/passivity of resistors.

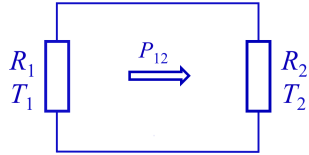

**Figure 1.** Energy transfer between two resistors. The warmer resistor is heating the other one. According to Equation 1, the warmer resistor is an active element while the colder one is a passive one. In reality, both resistors are passive.

## 2.2 Devices with idealized rectifier characteristics

As an example for a nonlinear device, we consider a hypothetical "idealized rectifier" (e.g. an idealized diode that is noise-free, etc.) which is a device that rectifies all signals, including arbitrarily small signals, such as the thermal noise of a resistor. For our studies, any current-voltage characteristics that are continuous asymmetric functions are satisfactory provided they can hypothetically extract a non-zero DC voltage from thermal noise.

When the rectification of thermal noise is ignored, the engineering definition of passivity/activity (based on Equation 1) yields the well-accepted conclusion that the idealized diode is a passive device. However, as it is shown in Section 4, in thermal equilibrium, feeding the hypothetical diode (and connected passive circuit elements) with thermal noise results in Equation 5, thus the diode circuitry is then an active device in this situation, according to the engineering definition. This paradox will be detailed and resolved in  Section 4.

### 3. The new, statistical physical definition of passivity/activity

Here we introduce our definition of passivity/activity based on the Second Law of Thermodynamics, that is, the impossibility of Perpetual Motion Machines (PMM) of the second kind. Such machines would be able to extract energy from the heat in thermal equilibrium without using extra energy for the process, that is, they would be able to cause steady-state temperature differences and decrease the entropy in a closed system in thermal equilibrium. That extracted energy by the PMM would violate also the Energy Conservation Law. The new definition below is self-consistent and provides robust definition of passivity/activity even in situations when the advanced engineering definition given in Section 1 fails to do so.

Note: the *device-in-question* below may need external energy sources (bias) to provide its required signal response. Example: a transistor requires a power supply for its bias conditions to operate. We will address this problem below.

Our definition of activity/passivity is:

a) The *device-in-question* is *active* if the following condition holds. Suppose a *hypothetical-device*, which does not require an external energy source to function and it has the same signal-response characteristics as the device-in-question. In an isolated system originally in thermal equilibrium, such a hypothetical-device in a proper circuitry would be able to produce steady-state entropy reduction in the system where all the other elements are passive. In other words, such hypothetical device would violate the Second Law of Thermodynamics.

Note: for example, such an entropy reduction could be a steady-state, non-zero temperature gradient in the device's environment; etc.; anything that can break thermal equilibrium conditions and persist over the duration of the operation of the device.

b) Physical implication: Such a hypothetical-device cannot exist in practice, thus an active physical device always requires an external energy source to execute its response characteristics of activity. *However, it is important to note that, if a device requires a power supply for its operation that does not imply that the device is an active one.*

c) The device-in-question is *passive* if it is not active.

### 4. Proof that idealized rectifiers would require active devices

While a real diode is a passive device, an idealized device that can rectify thermal noise always requires the usage of active device(s). We show that such a rectifier must be an active device not only according to our new definition in Section 2, but also by the former, engineering definition of passivity/activity based on Equation 1.

Without studying the problem of thermal noise, the old definitions [2] imply that an idealized diode is a *passive* device, which is the commonly accepted opinion. (The simpler and usual argumentation is that the diode does not have negative differential resistance region in its characteristics, which is a narrower version of the passivity/activity definition based on by Equation 1).

#### 4.1 The Brillouin paradox and the Second Law

If an *idealized* (noise-free, etc.) *rectifier* device functions at thermal equilibrium, it can provide nonzero DC voltage (or current) in a proper circuitry with thermal noise excitation, where only passive elements are connected to the rectifier. Hence, this becomes a PMM of the second kind and the Second Law is broken.

This observation, the Brillouin paradox [7,8], was first made out by Brillouin [7] for an idealized diode where he analyzed the nonlinearity up to the second component of the Taylor series. However, he used a Langevin approach to introduce the noise, which was shown to be invalid by van Kampen [8]. While many physicists agree that an effective passive rectification of thermal noise would result in a PMM, no general proof for arbitrary diode characteristics exists in the literature about this claim.

Below we show a simple circuit model and use scaling analysis to prove that any nonlinearity that provides nonzero voltage (or current) when driven by thermal noise in thermal equilibrium leads to the direct violation of the Second Law.

### 4.2 Proof that an idealized rectifier allows perpetual motion machines

Figure 2a shows a parallel resistance-capacitor (RC) circuit at temperature $T$. The circuitry acts as a first-order low-pass filter of the thermal noise. Equation 2 and linear response theory imply that the mean-square thermal noise voltage on the capacitor is [3,6]:

$$\left\langle U_{\mathrm{RC}}^2(t) \right\rangle = \frac{kT}{C} \ , \tag{6}$$

which follows also from Boltzmann's energy equipartition theorem [6]. The thermal noise voltage is a pure a.c. voltage, thus its mean value is zero:

$$\left\langle U_{\mathrm{RC}}(t) \right\rangle = 0 \ . \tag{7}$$

This situation radically changes when we shunt this circuit with an idealized diode that can rectify thermal noise. Figure 2b shows a parallel diode-resistor-capacitor (DRC) circuit at temperature $T$. Then the diode will asymmetrically damp the thermal noise voltage on the capacitor where the voltage can be written as:

$$U_{\mathrm{DRC}}(t) = U_{\mathrm{ac}}(t) + U_{\mathrm{dc}} \ , \tag{8}$$

where, $U_{\mathrm{ac}}(t)$ is the suppressed a.c. component and $U_{\mathrm{dc}}$ is the d.c. component due to the asymmetric damping.

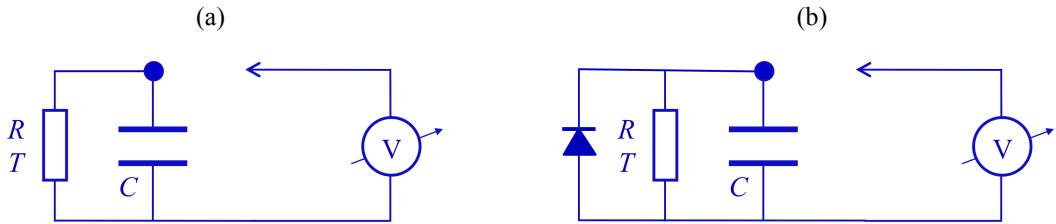

**Figure 2.** (a) A parallel resistance-capacitor (RC) circuit at temperature $T$. (b) A parallel diode-resistor-capacitor (DRC) circuit at temperature $T$.

The following relations naturally hold.

$$\left\langle U_{\mathrm{ac}}(t) \right\rangle = 0 \ , \tag{9}$$

$$\left\langle U_{\mathrm{ac}}^2(t) \right\rangle = \alpha \frac{kT}{C} \qquad , \text{where} \quad 0 < \alpha(T, R, C) < 1 \ , \tag{10}$$

$$U_{dc}^2 = \beta \frac{kT}{C} \qquad \text{, where } 0 < \beta(T, R, C) < 1 \; . \tag{11}$$

The actual value of $\alpha$ and $\beta$ are influenced also by the diode characteristics, and the (constant) values of $R$, $C$ and $T$ need not be specified, as they do not affect the conclusion.

Next, let us consider a serial ladder of parallel RC elements. Figure 3a shows an example of a ladder of $N$ serial RC units with $N$=2. Due to the statistical independence of thermal noise voltage in different RC units, the mean-square thermal noise voltage between the end contacts A and B is:

$$\langle U_{AB}^2(t) \rangle = N \langle U_{RC}^2(t) \rangle = N \frac{kT}{C} = \frac{kT}{C/N} \; , \tag{12}$$

that is, an equivalent situation as if, in Figure 2a and Equation 6, we would have used a capacitor with a reduced, $C/N$, capacitance.

Next, let us consider a serial ladder of parallel diode-resistor-capacitor (DRC) elements. Figure 3b shows a ladder of $N$ serial DRC units with $N$=2. The essential mechanism of the perpetual motion machine utilizes the fact that the AC components are statistically independent thus their mean-square value scales up by the factor $N$. On the other hand, the DC components are totally correlated, thus their square scales up with $N^2$. The mean-square AC voltage between electrodes A'B' is

$$\langle U_{A'B'ac}^2(t) \rangle = N\alpha \langle U_{RC}^2(t) \rangle = N\alpha \frac{kT}{C} \; , \tag{13}$$

$$U^2_{\text{A'B'dc}} = N^2 \beta \frac{kT}{C} \quad . \tag{14}$$

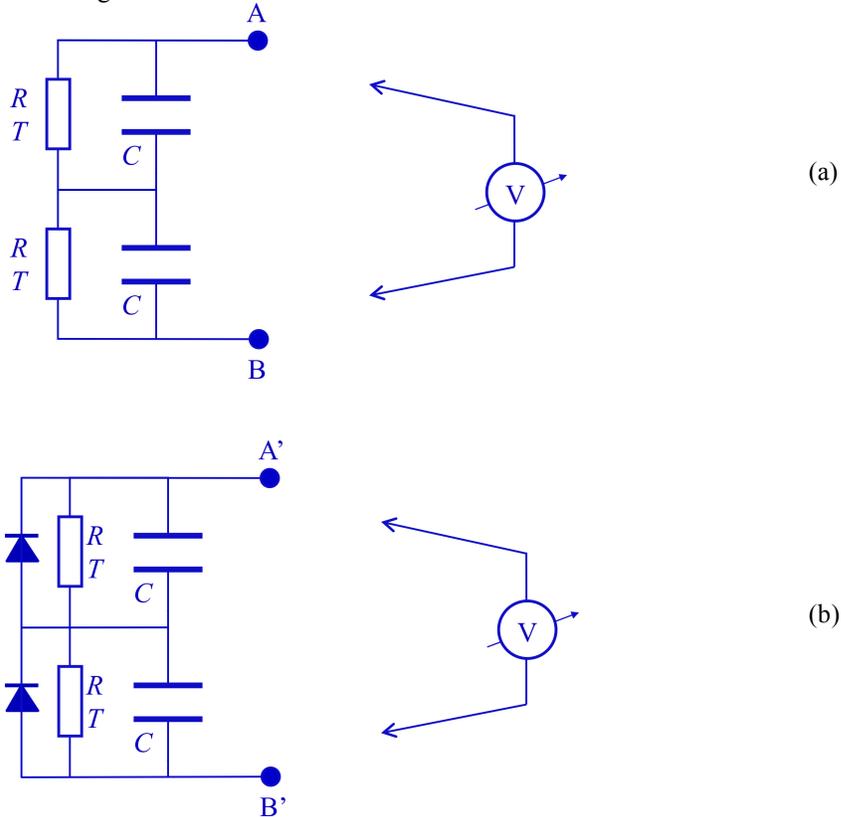

**Figure 3.** (a) A ladder of $N$ serial RC elements; in this example $N$=2. (b) A ladder of $N$ serial DRC elements, where $N$=2.

Finally, Figure 4a shows a modification of the circuit of Figure 3b after removing the resistors from the ladder circuitry and replacing them by a single external resistor of the same total value of *NR* (in the figure, *N*=2). If we suppose that the diodes are identical, the crosslinks between the diodes and the capacitors connect equipotential points. Thus, these crosslinks can be removed without any change in the circuit 's behavior. The equivalent circuit obtained by this modification is shown in Figure 4b for the case of *N*=2. We recognize that the circuits in Figures 2b and 4b are essentially the same as a single DRC circuit with the minor modification that, in Figure 4b, the resultant resistor and capacitor values are *NR* and *C/N*, respectively, and the diode characteristic is modified but not terminated because it is *N* serial diodes with identical polarity directions. Thus Equations 9-11 hold with different coefficients but the same limit.

$$\left\langle U_{\text{A''B''ac}}(t) \right\rangle = 0 \quad , \tag{15}$$

$$\left\langle U^2_{\text{A''B''ac}}(t) \right\rangle = \alpha_N \frac{kT}{C/N} = N\alpha_N \frac{kT}{C} \quad , \text{ where } \quad 0 < \alpha_N < 1 \quad , \tag{16}$$

$$U^2_{\text{A''B''dc}} = \beta_N \frac{kT}{C/N} = N\beta_N \frac{kT}{C} \quad , \text{ where } 0 < \beta_N < 1 \quad . \tag{17}$$

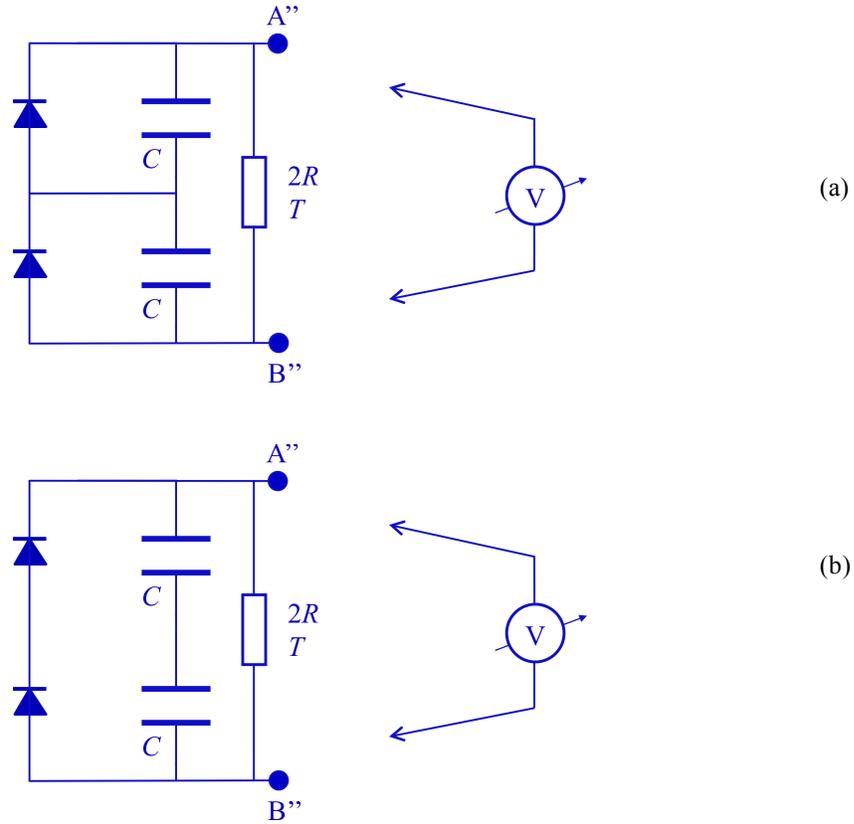

**Figure 4.** (a) The circuit in Figure 3b after removing the resistors from the ladder circuitry and replacing them by a single external resistor of value of *NR* (*N*=2). (b) The circuit equivalent to Figure 4a after removing the crosslinks between the equipotential points between the diodes and capacitors in the ladder circuitry (*N*=2).

Now, it is easy to see the fictional "energy source" of a perpetual motion machine. If $N$ is approaching infinity, the only term that scales with $N^2$ will dominate in the mean-square voltage terms. And only Equation 14 and the corresponding Figure 3b has this scaling. Thus, if we connect the contacts A' to A" and B' to B", respectively, there is a non-zero mean power flow from the circuitry in Figure 3b into the circuitries of Figure 4:

$$\lim_{N \to \infty} P_{\text{A'B'A"B"}} \propto \lim_{N \to \infty} \frac{U^2_{\text{A'B'dc}}}{NR} \propto \lim_{N \to \infty} N^2 \beta \frac{kT}{NRC} \propto \lim_{N \to \infty} N = \infty \quad . \tag{18}$$

Therefore, in the large $N$ limit, the heating power flow is proportional to $N$ even when the system starts from thermal equilibrium. That means the perpetual motion machines of the second kind can be created and the Second Law is violated.

*4.3 Activity of the hypothetical rectifier*

Equation 18 implies that both the engineering definition based on Equation 1 and our improved, statistical physical definition (see Section 3) of passivity/activity classifies the circuit in Figure 3b as active element. Moreover, our definition also implies that *any rectifier that can rectify thermal noise must be an active element*. However, a real diode driven by thermal noise only is a passive element and cannot rectify thermal noise.

# 5. Conclusions

We have shown that the advanced engineering definition of passivity/activity is self-contradictory when a thermodynamical system executing Johnson-Nyquist noise is present. A new, self-consistent definition based on the Second Law of Thermodynamics was introduced.

It was also shown that any rectifier circuitry that can rectify thermal noise with a noise temperature that is equal to the temperature of the rectifier must contain an active circuit element, according to both the engineering and statistical physical definitions.

N.G. van Kampen [9] showed that the particular mechanism via which the diode "avoids" rectification of thermal noise is an asymmetric thermal noise generated by itself the diode.

Finally, we note that, such an active device can be a diode itself provided it is driven by a proper external signal superimposed on thermal noise and the diode acts as an active parametric amplifier device. In this case, both the statistical physical and the engineering definitions may indicate an active device. However this topic is out of the scope of the present paper.